\documentclass[preprint,12pt]{elsarticle}

\usepackage{graphicx}
\usepackage{amssymb}
\usepackage{paralist}
\usepackage{array}
\usepackage{lineno}
\usepackage[utf8]{inputenc}
\usepackage{rotating}
\usepackage{booktabs}
\usepackage{multirow}
\usepackage{url}
\usepackage{color}

\journal{Advances in Space Research}

\begin{document}
\begin{frontmatter}

\title{Hesperos: A geophysical mission to Venus}

\author[1]{Robert-Jan Koopmans\corref{cor1}}
\ead{koopmans@fotec.at}
\author[2]{Agata Bia\l{}ek}
\author[3]{Anthony Donohoe}
\author[4]{María Fernández Jiménez}
\author[5]{Barbara Frasl}
\author[6]{Antonio Gurciullo}
\author[7]{Andreas Kleinschneider}
\author[8]{Anna \L{}osiak}
\author[9]{Thurid Mannel}
\author[10]{Iñigo Muñoz Elorza}
\author[11]{Daniel Nilsson}
\author[12]{Marta Oliveira}
\author[13]{Paul Magnus Sørensen-Clark}
\author[14]{Ryan Timoney}
\author[15]{Iris van Zelst}

\address[1]{FOTEC Forschungs- und Technologietransfer GmbH, Wiener Neustadt, Austria}
\address[2]{Space Research Centre Polish Academy of Sciences, Warsaw, Poland}
\address[3]{Maynooth University, Department of Experimental Physics, Maynooth, Ireland}
\address[4]{Universidad Carlos III, Madrid, Spain}
\address[5]{Zentralanstalt für Meteorologie und Geodynamik, Vienna, Austria}
\address[6]{Royal Institute of Technology, Stockholm, Sweden}
\address[7]{Delft University of Technology, Delft, The Netherlands}
\address[8]{Institute of Geological Sciences, Polish Academy of Science, Wroclaw, Poland}
\address[9]{Institute for Space Research, Austrian Academy of Sciences}
\address[10]{HE Space Operations GmbH, Bremen, Germany}
\address[11]{Luleå University of Technology, Luleå, Sweden}
\address[12]{Instituto Superior Técnico, Lisbon, Portugal}
\address[13]{University of Oslo, Oslo, Norway}
\address[14]{University of Glasgow, Glasgow, United Kingdom}
\address[15]{Institute of Geophysics, ETH Z\"urich, Z\"urich, Switzerland}

\cortext[cor1]{Corresponding author}

\begin{abstract}
The Hesperos mission proposed in this paper is a mission to Venus to investigate the interior structure and the current level of activity. The main questions to be answered with this mission are whether Venus has an internal structure and composition similar to Earth and if Venus is still tectonically active. To do so the mission will consist of two elements: an orbiter to investigate the interior and changes over longer periods of time and a balloon floating at an altitude between 40 and 60\~km to investigate the composition of the atmosphere. The mission will start with the deployment of the balloon which will operate for about 25 days. During this time the orbiter acts as a relay station for data communication with Earth. Once the balloon phase is finished the orbiter will perform surface and gravity gradient mapping over the course of 7 Venus days.
\end{abstract}


\end{frontmatter}


\section{Introduction}
\label{intro}
Since 1975, the Austrian \emph{Research Promotion Agency} (FFG), organises yearly the Alpbach Summer School. During the ten-day course, an intensive in-depth programme around a central space related theme is taught to about 60 graduates, post-graduates and young scientists and engineers from ESA member and cooperating states. The taught programme is complemented with a design exercise, where students propose and work out in groups a space mission related to the central theme. At the end of the course, all student groups present their mission to a panel of experts as well as to all other students, tutors and lecturers. The panel then chooses one or more proposed missions to be further worked out by a selection of students during the post-Alpbach week held in Graz.\\
The theme of the Alpbach Summerschool 2014 was "The geophysics of the terrestrial planets" and focussed, as the name suggests, on the inner planets, including the Moon, of the solar system. Independently, all four student teams in Alpbach decided to design a mission to Venus. Despite the fact that the same planet was chosen, four very different missions were presented at the end of the summer school. Two of these missions were combined to one and worked out further during a further week, called the post-Alpbach week.\\
In this paper the result from the post-Alpbach week is presented. The aim of the mission to Venus is to investigate the tectonic activity and its associated time scale as well as study its core and mantle. The mission is named Hesperos, which refers to the evening star or planet Venus in the evening in Greek mythology. Hesperos was thought to be the brother of Phosphorus: the morning star or planet Venus as seen in the morning. It was only realised later that the two stars represented the same celestial body.\\
This paper starts with a short review of past missions to Venus and what has been learned from these missions in context of the proposed mission. This forms the starting point of formulating the science objectives in the section that follows. These science objectives are then translated into observables. In the subsequent section the payload is presented that is required to fulfil the science objectives. The required spacecraft design is discussed next with special emphasis on those aspects that make a mission to Venus challenging. In the section after that the mission timeline is discussed and broken down in several phases. Finally, conclusions are drawn about a couple of key aspects of the mission.\\

\section{Past Missions}
\label{past}
Recently, strong evidence was presented for present-day active volcanism on Venus in the form of local, transient bright spots measured by the Venus Express Monitoring Camera \cite{shmb15}. The transient spots were interpreted as an elevated temperature due to extrusive lava flows. They were identified in the relatively young Ganiki Chasma and are similar to locations on Earth that are associated with rift volcanism. This new finding strengthens the hypothesis that Venus is geologically as well as geodynamically active today. \\ 
Active volcanism could be confirmed by the Akatsuki mission, which was inserted in an orbit around Venus in 2015 after a previous attempt failed \cite{nakamura2014}. Finding evidence of active volcanism is an important secondary objective of Akatsuki, although the main mission objectives are to study atmospheric dynamics and cloud physics. \\
Volcanic activity on Venus could also explain the temporal variations in atmospheric SO$_2$ measured by Pioneer Venus (e.g., \cite{esposito1984}) and Venus Express \cite{mabm13}. Pioneer Venus performed in situ measurements in the atmosphere to determine, amongst others, its chemical composition. For this purpose, Pioneer Venus released four atmospheric entry probes, one large one and three smaller ones. The spatial variation of SO$_2$ in the Venusian troposphere was measured by the VEGA probes \cite{bertaux1996}. In addition to the probes measuring the chemistry of the atmosphere, both VEGA satellites made use of a meteorological balloon with an instrumented gondola to study the temporal behaviour and horizontal structure of the Venus atmosphere \cite{salb86}. \\
Volcanism itself is the surface expression of processes taking place in the interior of the planet of which very little is known. Other than that Venus has a distinct crust, as was concluded from the Venera 8 mission \cite{visk73}, there is still a debate about the size and phase of the core.\\
Pioneer Venus and Magellan were missions that contributed to a better understanding of the interior of Venus by constructing topographical maps, measuring surface characteristics and mapping the gravity field. The Pioneer Venus Orbiter mapped 93\% of the surface of Venus with a surface resolution better than 150\,km \cite{peef80}. The vertical accuracy was typically 200\,m \cite{peef80}. NASA's Magellan spacecraft was mainly dedicated to map the surface of Venus with a synthetic aperture radar and to measure the gravity field. The horizontal resolution of the radar imagery varied between 120 and 300\,m, depending on the altitude of the spacecraft \citep{pefj91}. The accuracy of the measurements varied between 50 and 100\,m, with a maximum accuracy of 5 m for individual measurements with a horizontal resolution of 10 km  \citep{FordPenttengill1992}. A variety of deformational features was imaged by Magellan, such as families of graben, wrinkle ridges, ridge belts, mountain belts, quasi-circular coronae and broad rises with linear rift zones with dimensions of hundreds to thousands of kilometers \cite{sosb92}. These last two deformation types are not observed on Earth and appear to be unique for Venus.\\

\section{Science Objectives and Observables}
\label{science}
Although Venus and Earth are similar in size, mass and distance from the Sun, the chemical composition of the atmosphere, surface pressure and temperature and rotation reveal that they are very different worlds. The reason for these differences are not understood. The aim of the Hesperos mission is to gain insight in why and how Earth and Venus evolved so differently. For this purpose the geophysics of Venus is investigated by exploring the following scientific questions:
\begin{inparaenum}[1.]
\item Does Venus have a similar internal structure and composition as Earth? \label{SciObj1}
\item Is Venus tectonically active and on what time scale? \label{SciObj2}
\end{inparaenum}
Answering these questions will not only refine evolution models of Venus but also constrain planetary evolution models in general and conditions necessary for emergence of life on our planet as well as on others, including exoplanets.\\
The importance of this scientific topic has been recognized in ESA’s and NASA’s strategic plans. To answer the first question stated by the Cosmic Vision (ESA, 2005) it is necessary to study what conditions are necessary for planet formation and emergence of life. Similar questions are raised in Visions and Voyages (National Research Council, 2011) in the Building New Worlds section. In the following, each scientific question will be elaborated on and supplemented by formulating sub-questions.

\subsection{Internal structure and composition}
\label{subSciObj1}
While the internal structure of Earth is relatively well known due to seismological studies (e.g., \cite{dzan81}), the interior of Venus is not well constrained. At present, the internal structure of Venus is often assumed to be similar to that of Earth, because of the comparable mass, radius and density. However, direct evidence for this assumption is not available. Therefore, the first science objective of the Hesperos mission is to provide more constraints on the internal structure and composition of Venus. To help achieve this objective, two sub-questions have been formulated:
\begin{inparaenum}[(a)]
\item What is the size of the core and its phase?
\item How do mantle processes drive surface activity?
\end{inparaenum}

\paragraph{Core size and phase}
Estimates of the moment of inertia factor, which characterises the radial mass distribution in a planet, vary widely for Venus. Consequently, estimates of the size of the core of Venus have large uncertainties. From Doppler tracking data of the Magellan and Pioneer Venus Orbiter missions, the second harmonic potential $k_2$, or Love number, has been estimated to be $k_2 = 0.295\pm0.066$ \cite{koyo96}. Models predict a $k_2$ value of $0.23 \leq k_2 \leq 0.29$ for a liquid iron core and $k_2 = 0.17$ for a solidified iron core \cite{koyo96}. Hence, the Doppler data suggests that Venus' core is liquid. However, these studies implicitly assumed that the viscosity of Venus' core is similar to Earth's \cite{bohp97}.\\
To estimate the moment of inertia factor of Venus and extrapolate from this the core size, variations in the spin state of the planet have to be determined accurately. We focus on measuring the spin rate of Venus, which has already been determined several times in the past. However, the measured values and their corresponding uncertainties are significantly different from each other \cite{muhp12} as shown in Table~\ref{spin_rate}. The Magellan mission measured the spin rate with an uncertainty of $\pm0.0001$, which amounts to a precision of about 2100\,s. For meaningful results from which the interior of Venus can be better constrained, the length of a Venus day has to be measured with a precision of 10\,s or at least 200 times better than the current available data \cite{cottereau2011}.

\begin{table}[ht]
\setlength\extrarowheight{2pt}
\centering
\caption{Measurements of the spin rate of Venus, modified from \cite{muhp12}. For additional references, the reader is referred to \cite{muhp12}}. \label{spin_rate}
\begin{tabular}{ll} \hline
\textbf{Observations} & \textbf{Period of rotation (days)} \\ \hline
Goldstone 1972–1982 & 	243.022 $\pm 0.003$	\\ 
Earth based 1972–1988 &	243.022 $\pm 0.002$	\\ 
Earth based	& 243.025 $\pm 0.002$	\\ 
Magellan gravimetry &	243.0200 $\pm 0.0002$	\\ 
Magellan SAR &	243.0185	$\pm 0.0001$\\ 
Venera \& Magellan SAR & 243.023 $\pm 0.001$	\\ 
Earth based 1975–1983 & 243.026 $\pm 0.006$ \\ \hline
\end{tabular}
\end{table}

A way to constrain the thermal evolution and core evolution of Venus is to study the magnetic field of Venus. If Venus has a magnetic field or if remnants of a magnetic field can be detected on Venus, (numerical) models of Venus' thermal evolution and core growth can be used to determine the most likely phase of the core at present. So far, a magnetic field has not been measured on Venus. This is due to the difficulty of detecting the intrinsic magnetic field of Venus in the presence of the induced magnetosphere \cite{luhmann2015}. The results from the Pioneer Venus Orbiter are consistent with a ~3\,nT equatorial dipole magnetosphere model, which is $10^4$ weaker than the magnetic field of the Earth.

\paragraph{Mantle processes}
As the radius of Venus is known, constraints on the size of the core also provide constrains on the size of the mantle. However, this does not reveal anything about the structure or dynamics of the mantle. Mantle dynamics and structure are, at least on Earth, closely linked to activity on the surface. As such, knowledge of Venus' mantle is essential in understanding its surface features. \\
Studying the structure of the mantle from orbit mainly involves studying the mass distribution within the mantle. The density in the mantle varies, for example, when a hot plume causes an expansion of the mantle material which results in a decrease in density. These local variations in density result in local differences in the gravity gradient \cite{papg14}. The main advantage of measuring the gravity gradient over the simpler and more conventional gravity mapping, is that the gravity gradient is more localised to the source of the anomaly. This makes this method more suitable for linking the gravity gradient anomalies in the mantle to their surface expressions. Surface activity investigations are therefore a natural extension and are discussed in greater detail in the next section. Mapping of the gravity gradient to gain insight into mantle dynamics has already been successfully applied to Earth by using data from the GOCE mission \cite{papg14} and the authors speculate that this new method can contribute to constraints of the mantle density and viscosity on global and regional scales on Earth. GOCE measured six components of the Earth's gravity gradient tensor, of which four were determined with an optimal accuracy of 1-2 $\cdot 10^{-11}$s$^{-2}$ that was reached at scales smaller than 750\,km \cite{papg14}. 

\subsection{Tectonic activity and time scale}
\label{subSciObj2}
Plate tectonics play a major role in recycling chemicals essential for life, increasing atmospheric pressure by degassing and creating diverse environments where organisms can live \cite{Valenciaetal.2007}. It has even been argued that life as we know it could not have arisen without plate tectonics. Mercury, Mars and the mMoon were tectonically active in the past, but this activity was not related to plate tectonics \cite{Harrison2000}. The only terrestrial planet in the solar system next to Earth that may still be volcanically active and may have plate tectonics is Venus.\\
To investigate this, the second scientific question is divided into two sub-questions:
\begin{inparaenum}[(a)]
\item Is there plate movement on Venus and what are its characteristics?
\item Is there volcanic activity on Venus and to what extent?
\end{inparaenum}

\paragraph{Plate movement}
There are currently several contrasting theories of Venus’ tectonics. The `stagnant lid theory' \cite{Turcotte1993,BasilevskyHead2000} states that no terrestrial-type plate tectonics occurs. Instead, heat accumulates in Venus’ mantle. This leads to episodic catastrophic resurfacing that most recently happened $\sim$500 Ma ago (e.g.,  \cite{scsm92} and \cite{Fegley1995}). During this time a large part of the surface was covered by thick lava flows. This theory assumes that there are no tectonic plates on Venus and surface activity between those catastrophic events is limited. Other theories assume that plate tectonic regimes analogous to that of Earth exist on Venus. For example, \cite{SchubertSandwell1995} suggest that subduction and rifting processes similar to those on Earth could account for the thermal evolution of Venus and could potentially explain the observations of the northern and southern margins of the Latona Corona as subduction sites. A lack of active volcanism in this latter region could corroborate this theory. \cite{ghail2015} suggests subcrustal lid rejuvenation as an analogue to Earth's plate tectonics to account for the observed largely stable tectonic regime with subcrustal horizontal extension. The best opportunity to detect and measure these movements of Venus' crust is to fly a mission capable of detecting change, for instance by using a radar, according to \cite{ghail2015}. Another end-member theory by \cite{JohnsonRichards2003} emphasizes the importance of (small-scale) mantle plumes to explain Venus' tectonic regime. \cite{JohnsonRichards2003} mainly provide a mechanism for the formation of coronae, namely small-scale mantle upwellings.  They hypothesized that the coronae in the lowland and plain regions are older on average than those in the Beta, Atla, and Themis Regiones, because of the interference beneath the plains of mantle downwellings with small-scale upwellings. Observations that could support this theory include the observation of partially buried coronae due to volcanism and global mapping of coronae to deduce their relative ages. \\
The most obvious way of investigating tectonics from orbit is by determining the thickness of the crust and combine this information with detailed topographical maps. Obtaining this data allows to detect subsurface structures such as rift systems and subduction zones. Variations in the crustal thickness result in uneven mass distributions in the upper mantle \cite{papg14} and the lithosphere \cite{Boumanetal2013}. This causes small variations in the gravitational field. On Earth, such structures are associated to gravity anomalies of a magnitude of 10-30\,mgal and become visible at spatial resolutions of 50-100\,km.\\
Rift-like features have been observed on Venus, and they are of similar size as terrestrial ones. They are thousands of kilometres long and tens of kilometres wide \cite{Ghail2002}. Unfortunately, current resolution of Venus’ gravity field is known at a spatial resolution of 700\,km \cite{Konoplivetal1999}. This does not allow to distinguish those features in the gravity field. In order to detect rifts and possible subduction (and/or obduction zones \cite{Ghail2002}) changes in gravitational field of 5\,mgal and with spatial resolution $<$100\,km need to be detected.\\
Current global topographical maps obtained with Magellan have a horizontal resolution of 10-20\,km and vertical resolution of 5-100\,m \cite{FordPenttengill1992}. About 20\% of the planet’s topography is known with a spatial resolution of 1-2\,km and a vertical accuracy of 50\,m \cite{Herricketal2012}. To achieve the science objective neither of these is sufficient. The minimum required spatial resolution is 40\,m and vertical accuracy on the order of meters.\\
One of the most evident manifestations of tectonic activity is an earthquake. Usually earthquakes are studied with seismometers deployed carefully on the surface \cite{Khanetal2013}. However, the harsh conditions on the surface of Venus make this kind of investigation less feasible. However, it is known that due to seismic coupling with the atmosphere quakes can be detected in the atmosphere \cite{Kanamorietal1994}. Acoustic waves generated by Rayleigh surface waves have been observed during earthquakes \cite{Artruetal2004,Dautermannetal2009}. On Venus, quakes with the same magnitude are predicted to generate infrasound waves with amplitudes 600 times larger than recorded on Earth at the same height \cite{Garciaetal2005}. This is due to the different atmospheric pressure and density. A difficulty is that the background noise in the atmosphere at Venus is largely unknown. Especially in the mid-cloud region the atmosphere is particularly turbulent. To measure quakes in the atmosphere, the pressure should be measured at two different locations with a vertical separation from each other of a couple of tens of meters. By comparing the signals one can discriminate between pressure waves from seismic origin and pressure waves generated by other mechanisms~\cite{KISS2015}.\\ 
Another important feature of tectonic activity is degassing of the interior. The extent of degassing can be determined from measurements of the ratios of radiogenic noble gases in the atmosphere. Measurements of $^{40}$Ar/$^{36}$Ar ratio, performed by Venera 11 and 12, gave a value of 1.19 +/- 0.07 and were confirmed by Pioneer Venus to be 1.03 $\pm$0.04 \cite{Fegley1995}. Those results indicate that  Venus is much less degassed than Earth, Mars or even Titan, where the ratios are in the range of 150-2000 \cite{Owen1992,Atreyaetal2006,Pujoletal2013}. The low degassing rate on Venus has strong implications for the thermal and tectonic evolution of Venus and is in favour of the stagnant lid theory \cite{SolomatovMoresi1996}. However, since lower amounts of $^{40}$Ar can also be explained by significantly lower amounts of $^{40}$K, from which radiogenic Argon is produced, an independent isotope ratio, such as $^3$He/$^4$He, needs to be measured as well to verify the extent of degassing of Venus. To do this an instrument is required capable of measuring with a resolution of 0.1\,ppb and a resolution of 0.1\,AMU at least. The measurement range should be in the range of 2 to 45\,AMU.

\paragraph{Active volcanism}
Volcanic activity is one of the most prominent manifestations of the internal activity of a planet. Multiple lines of evidence suggest that Venus is a volcanically active planet. Venus is similar in size to Earth and its cooling rate should therefore be sufficient slow to still support volcanism. Geochemical composition of relatively fresh rocks measured by Venera landers is consistent with volcanic rocks \cite{Surkovetal1983}. Their age is not known, but dating performed by crater counting shows that the surface of Venus is young, likely less than 500\,Ma \cite{McKinnonetal1997}. Besides that it is covered by numerous landforms that resemble volcanoes \cite{BasilevskyHead1998}. In addition, the measured variation of the atmospheric abundance of SO$_2$ has been interpreted to be a result of volcanic activity \cite{esposito1984,Marcqetal2013}. However, other explanations such as long term variation in the circulation within the atmosphere \cite{ClancyMuhleman1991} are also possible. Recently, short term heat pulses, in the order of a few days, were detected on the surface by the Venus Monitoring Camera on board of Venus Express \cite{shmb15}. They were interpreted as magma or hot volcanic release and thus a strong indication of volcanic activity. However the extent of this has not yet been determined.\\
Determining the extent of active volcanism on Venus can be performed by investigating the different manifestations of volcanism. The most obvious one is the heat signature. Lava from active volcanoes produces a clear signal in the IR spectrum. 
The spectral range is governed by Venus' atmospheric window, which only allows certain wavelengths (in the range 0.8 - 1.2\,$\mu$m \cite{taylor1997}) to pass through the optically thick clouds.\\
Lava deposition results in small changes in the morphology of the surface. The extent of these changes is dependent on the size of the volcano. Morphology changes can be detected by comparing radar images of the surface from different moments in time. Previous missions such as Magellan have already conducted radar surveys whose results can be used as reference. Should activity be observed it is possible to determine the level of activity by repeating measurements. A resolution $\leq$ 50\,km is sufficient to resolve most shield volcanoes and all pancake volcanoes as well as pyroclastic flows and impact craters. A volcanic eruption that would be rated as VEI (volcanic explosivity index, \cite{Newhall1982}) 3 or higher would deposit enough material that a change in topography can be resolved with a vertical resolution of 25\,m. An instrument that meets these requirements should be able to detect new volcanic activity.\\
Another indication of volcanic activity can be obtained by monitoring volcanic gases such as SO$_2$, H$_2$O and HCl abundance variations in the atmosphere over time. Measurements of SO$_2$ can be conducted in the spectral range of 0.2 to 0.3\,$\mu$m, which fits well in the Venus' spectral window. However, the mentioned volcanic gasses do not necessarily originate from volcanic activity. Therefore, measuring isotopic patterns of sulphurous gases provides constrains to the identification of potential sources for those gases and their circulation patterns. The light stable isotopes, in this particular case hydrogen (H, D), oxygen ($^{16}$O, $^{18}$O) and sulphur($^{32}$S, $^{34}$S), provide information on composition of the volcanic gases, their distribution in the atmosphere, the source material, with implications for planet formation, as well as isotope fractionation processes within the atmosphere of Venus. One advantage measuring isotopes of the volcanic gases in addition to their abundance is that mass-dependent isotope fractionation is temperature sensitive and isotopic distribution offer more detailed information into the processes involved. Ideally, measurements are performed \textit{in-situ} at different altitudes during extended periods of time. Especially  measurements within the cloud layer are particularly interesting in order to test a hypothesis of sulphur circulation in the mesosphere \cite{ClancyMuhleman1991}.\\
Finally, if volcanic eruptions currently occur on Venus, volcanic ash may be present in the higher atmosphere. Volcanic ash is composed of fragmented volcanic glass and pulverized rock \cite{Bukowieckietal2011}. Under terrestrial conditions volcanic ash has a relatively short residence time in the troposphere - and most of the particles fall back on the surface within a day at most. The smallest particles, very fine ash $<$10\,$\mu$m in diameter, can be transported to the troposphere or lower stratosphere where they can stay there for several days. Volume fraction of grain size generally decreases with size in the fine ash range. In order to capture the grain size distribution of those airborne particles measurements in the range from 0.1-10\,$\mu$m and with accuracy of at least 0.1\,$\mu$m are required.\\
Detecting volcanic ash in the atmosphere, complemented with the measurements of volcanic gas abundances and isotopes could give strong evidence for proof of very recent and/or current volcanic activity.

\section{Payload}
\label{payload}
As has become clear from the previous section, a wide range of different parameters have to be measured to fulfil the mission objectives. Some of the parameters can be measured by the same type of instrument while for others different types are required. 
Table~\ref{observables} summarises the required observables for each sub-question that has been discussed in the previous section. The last column indicates for each type of observable the required instrument type to perform the measurements with. They will be discussed in more detail below.

\begin{table}[h!t]
\setlength\extrarowheight{2pt}
\centering
\caption{Summary of observables. (O): orbiter, (B): balloon}
\label{observables}
\begin{tabular}{lll} \hline
\textbf{Objective} & \textbf{Observable} & \textbf{Instrument type} \\ \hline
\multirow{2}{2.5cm}{Core size and phase}
 & - spin rate & radar (O) \\
 & - magnetic field & magnetometer (B) \\ \hline
\multirow{2}{2.5cm}{Mantle processes}
 & - gravity gradient & gravity gradiometer (O) \\
 & - topography & radar (O) \\ \hline
\multirow{4}{2.5cm}{Plate movement}
 & - gravity gradient & gravity gradiometer (O) \\
 & - topography & radar (O) \\
 & - acoustic waves & infrasound detecto (B) \\
 & - isotopic ratios of noble gases & NMS/TLS (B) \\ \hline
\multirow{5}{2.5cm}{Active volcanism}
 & - topography & radar (O) \\
 & - surface heat signature & spectrometer/camera (O) \\
 & - amount of volcanic gases & spectrometer (O) \\
 & - isotopic ratios of volcanic gases & NMS/TLS (B) \\
 & - amount of volcanic ash & nephelometer (B) \\ \hline
\end{tabular}
\end{table}

\paragraph{Radar}
A synthetic aperture radar (SAR) will be used to measure the surface topography and determine the spin rate of the planet. The resolution of a radar scan is determined by the length of the synthetic aperture \cite{Doerry2004}, which can be adjusted by varying the frequency of the pulses, therefore it is possible for one radar to have several operational modes with differing resolution and therefore differing data-rate. For the primary objective of generating detailed topographical maps, a resolution of $\leq$ 50\,km is sufficient for the identification of volcanic features. As will become clear from section~\ref{orbSciPhase}, the SAR will only be operated during part of the orbit. The resulting data rate is therefore low enough to send to the Earth.\\
In measuring the spin rate, a reference point on the surface of Venus is required. The Venera landers, whose remains are still present on the surface of Venus and whose location is roughly known, are large and made of metal which is highly reflective to radar. This makes them a perfect candidate for reference points on the surface of Venus. Although no longer in operation, their remains will still be visible to a radar with a 20 \,m resolution. By measuring the position of the landers several times during the mission lifetime the variation of the spin rate over time can be determined. Only for those parts of the surface where the Venera landers are, the resolution of the SAR will be temporarily increased to 20 \,m. Alternatively, spin rate measurements can be performed by position measurement of surface features as well, as was done by Mueller et al.\,\cite{muhp12}.

\paragraph{Gravity gradiometer}
Gravity gradient measurements will be performed with a gravity gradiometer. As was mentioned in section~\ref{subSciObj2} this instrument needs to have a spatial resolution of $<$100\,km and be able to measure changes as low as 5\,mgal. This can be obtained with a cold atom gradiometer on an orbiter \cite{Carrazetal2014}. An important requirement for operating this instrument is that it is in a vibration free environment.\\
As a concept this atomic quantum sensor is beneficial due to the fact that it is not limited in the same ways as ordinary electrostatic gravity gradiometers. The new method have been proven to work reliable on Earth and, although it has not been used on any previous space mission, zero-g testing indicates the benefits of this technique. Reliable specifications for this instrument will be obtained after more thorough testing and development \cite{Muntinga2013}.

\paragraph{Spectrometer/camera - in orbit}
Spectrometers are required for the determination of the amount and isotopic ratios of noble and volcanic gases and a camera to investigate the heat signatures on the surface of the planet. Because of the wide range of requirements for the different measurements, several spectrometers will be employed, each suitable for a different spectral range.\\
An IR spectrometer and camera will be used for detecting spots with high thermal flux on the surface. This spectrometer is based on elements of the SPICAV and VIRTIS instruments on board Venus Express. To increase the chances of detecting hot spots the instrument should be operated as long as possible over the whole mission lifetime. Alongside the IR spectrometer an UV spectrometer will be employed similar to the SPICAV instrument on board Venus Express. 

\paragraph{Spectrometer/camera - in-situ}
In order to determine, more exactly, the amount of each chemical in the atmosphere as well as the isotopic ratios, a neutral mass spectrometer (NMS) and a tunable laser spectrometer (TLS) will be employed. The NMS includes a quadrupole mass analyser and a gas processing system which will be able to determine major and trace gas species with a high sensitivity (10$^{-6}$)~\cite{Brinckerhoff2010}. The TLS uses infra-red lasers to analyse air samples in a so called multi-pass Herriot Cell with a senistivity of 2\,ppm for water. The unique signature of the absorbed laser light provides a measure of the concentration of various gas molecule species (in this case SO$_2$ and H$_2$O) and their isotopes. A TLS as part of a sample analysis unit was already used on multiple planetary missions, like the SAM (Sample Analysis at Mars unit) on Curiosity~\cite{Webster2008}.\\
The NMS and TLS require to take samples of the atmospheric gases and need therefore be operated \textit{in-situ}. These instruments need to perform multiple measurement cycles with a required measurement accuracy of 5-10\,\% and 1-2\,\% for SO$_2$ and H$_2$O and their isotopes, respectively, as well as 5-10\,\% for noble gases (argon and helium isotopes). As there is a chance that the atmosphere contains solid particles the inlet of both instruments shall have an air filter to prevent solid particles from entering the instrument and clog and/or damage the system. Furthermore, ideally both spectrometers will sample at different altitudes to obtain a broader variability with along an atmospheric vertical profile and sample different atmospheric layers and their chemical composition.

\paragraph{Nephelometer}
A nephelometer will be used to measure the size and amount of volcanic ash/dust as well as the amount of H$_2$SO$_4$ in the atmosphere. The instrument takes gas samples from the atmosphere and illuminates them with a laser. The scattering of light due to the particles in suspension, i.e. ashes and aerosols, in particular sulphuric acid, allows to compute for size, shape refractive indices and composition \cite{Banfield2004}. A nephelometer matching the requirements mentioned in section~\ref{SciObj2} is presented in the Venus Climate Mission~\cite{VCM2010} and the Venus Flagship Mission~\cite{VFM2010}.\\
The nephelometer requires sampling of the atmosphere and need therefore be operated \textit{in-situ}. It is expected that ash is to be found in the lowest cloud layer at about 48 to 50\,km above the surface \cite{taylor2014}. 

\paragraph{Infrasound detector}
Earthquakes can be monitored by recording and cross-correlating signals from microbarometers~\cite{Mutschlecner2005}. Infrasound Monitoring Stations (IMS), conceived to detect nuclear explosions, often record pressure waves originating from Earthquakes, but also from, for instance, volcanoes~\cite{Dabrowa2011}. Monitoring quakes like this has the great advantage that a challenging landing mission is not required. However, a particular difficulty is that the environmental noise in the atmosphere of Venus is not known. Wind, pressure and temperature variations as well as turbulence all result in pressure fluctuations interfering with pressure oscillations resulting from quakes. Besides that, also the platform from which the measurements are taken gives rise to noise~\cite{KISS2015}.\\
To reduce the noise from the platform the microbarometers should be placed far away from it. This can easily be achiever by means of a tether. As was discussed earlier, to distinguish between pressure waves originating from quakes and those from other sources, two microbarometers with a vertical separation of a couple of tens of meters should be employed. With currently available microbarometers measurements in the range from 0.01 to 1\,Hz and 1 to 5\,Hz are possible.\\
It is estimated that for COTS microbarometers deployed in the Venusian atmosphere, quakes with a magnitude of 6 can be detected from a distance of 2200\,km and those of magnitude 7 from about 9000\,km. In case the platform happens to be above the epicentre of a quake, quakes with a magnitude down to 3 can easily be detected~\cite{KISS2015}. Note, that one of the objectives of the Hesperos mission is to detect quakes. This is in stark contrast with the study described in~\cite{KISS2015}, which aims at using seismic activity to probe the planet's interior structure. The payload proposed for the Hesperos mission is similar to the payload for the generation 1 missions, i.e.\, pathfinder missions, proposed in~\cite{KISS2015}. It that report it was concluded that the required technology for such missions is already available.

\paragraph{Magnetometer}
Up until now no magnetic field intrinsic to Venus has been detected. That means that the magnetometer has to be flown closer to the surface and/or be more sensitive. For any meaningful measurements the altitude at which the measurements should take place is bounded by the ionosphere. In the ionosphere and above the interaction of the solar wind with the atmosphere of Venus induces a strong, varying magnetic field~\cite{zhang2016}.\\
Two triaxial fluxgate magnetometers of the type of the Magnetospheric Multiscale Magnetometers flying on the Magnetospheric Multiscale (MMS) mission~\cite{Russell2014} shall be used. Their dynamic range lies between $\pm$650\,nT (low range) or $\pm$10.500\,nT (high range) with  a noise density at 1 Hz of less than 8\,pT/$\sqrt{Hz}$ (low range) or 100\,pT/$\sqrt{Hz}$ (high range).\\
To increase the accuracy of the measurements the magnetometers will be flown as close as possible to the surface. As every electric current will introduce a magnetic interfering field the duty cycle of the magnetometers must alternate with those of the other instruments on the balloon and the magnetometers will be placed on a boom. The length of the boom depends on the magnetic cleanliness expected for the balloon gondola which will be better than for a usual spacecraft due to the absence of orbital control units and solar panels.
To further decrease the influence of disturbances a technique already carried out during the Venus Express mission is adopted~\cite{pope2011}, namely to position one magnetometer at 2/3 of a boom to measure the magnetic background. This magnetic background is then subtracted from the measurements taken at the end of the boom.

\section{Mission Design}
\label{design}
As has become clear from the previous section, there is a wide variation in required altitude for the different instruments. The most obvious variation is that instruments such as the radar and gradiometer require an orbit with as little interference from the atmosphere as possible, while instruments such as the nephelometer and sounding device require positioning in the cloud layers of Venus. For this reason, the mission design consists of two components: an orbiter operating outside the atmosphere of Venus and a balloon performing measurements in the atmosphere of Venus. An overview of which instrument is carried by the orbiter and balloon is given in table~\ref{tab:payloadAlloc}. The estimated mass and power consumption of the payload is 271\,kg and 551\,W for the orbiter and 16\,kg and 69\,W for the balloon.

\begin{table}[ht]
\setlength\extrarowheight{2pt}
\centering
\caption{Payload for the orbiter and balloon.}
\label{tab:payloadAlloc}
\begin{tabular}{l l}
\hline
\textbf{orbiter} & \textbf{balloon}\\
\hline
SAR & NMS \\
cold atom gradiometer & TLS \\
IR spectrometer & nephelometer \\
UV spectrometer & infrasound detector \\
 & magnetometer \\
\hline
\end{tabular}
\end{table}

\subsection{Orbiter}
\label{orbiter}
The spacecraft design, and in particular the orbiter design, is driven by the high solar flux at Venus, the high required power and data rate for the SAR and the fact that the orbiter acts as a data relay station for the data returned by the balloon. Consequently, the thermal control, power and communications subsystems require special attention.\\
Thermal control is a considerable challenge for spacecraft operating around Venus. The sun's radiation (2.6\,kW/m$^2$), planetary albedo ($\sim0.8$), planetary IR radiation (0.15\,kW/m$^2$) \citep{VenusFactSheet} and internal power dissipation all contribute to heat load on the thermal control system. The orbiter uses a combination of multi-layered insulation (MLI), heater lines (during eclipse), louvers and radiating/reflecting surfaces for thermal control. The worst-case incoming heat flux is about 2.8\,kW/m$^2$. The heat load is rejected to space using two radiators of 8\,m$^2$ each. External MLI is composed of 23 layers of Kapton. High-reflectivity coating, heritage from CryoSat-2, is used on exposed surfaces. During eclipse, 16 redundant heater lines retain the temperature within the operational limits of the equipment and instruments.\\
The spacecraft's power is provided by solar arrays made of Gallium Arsenide cells, heritage of Venus Express. Based on the mission operation phases, as will be discussed in section~\ref{timeline}, the maximum expected power consumption is 1.3\,kW, including margins. A simple simulation was performed to estimate the power generated as a function of the position in orbit with respect to the sun. 
Conservative estimates of the solar cell efficiency (0.2) and coverage ratio (0.85) result in a required solar panel area of about 3\,m$^2$. During eclipse, the spacecraft runs on batteries recharged in sunlit parts of the orbit. The surface area is such that enough power is generated to charge the batteries.\\
%
%
The orbiter provides a link with ground stations on Earth. It also acts as a relay for the data generated by the balloon. Note that the balloon is only active in the first science phase of the mission, see section~\ref{timeline}. For the remaining phases no relay function is required.\\
Telemetry and telecommand are supported by three antennas on the spacecraft. The science data is returned via a High-Gain Antenna (HGA) in X-band; a typical choice for deep-space ESA missions. As the scientific instruments, especially the SAR, produce a considerable data volume a large steerable 3\,m dish antenna is chosen for the HGA. Note that Ka-band might be well established by the time a Venus mission enters the detailed design phase, resulting in an increased data rate and decreased antenna diameter, at the cost of a more demanding requirement for precise pointing. Supplementary to the HGA, an S-band Medium-Gain Antenna (MGA) provides a link to the ground stations when the spacecraft is close to Earth, during transfer and during safe mode and other occurrences when the HGA cannot be pointed accurately and no high data rate is required. Lastly, an UHF link keeps contact with the balloon. As ground stations the 35\,m dishes of ESA's Deep Space Antennas (DSA) are selected. An overview of the communication system elements is shown in Table \ref{tab:orbiter_comm}.

\begin{table}[ht]
\setlength\extrarowheight{2pt}
\centering
\caption{Elements of the communication system.}
\label{tab:orbiter_comm}
\begin{tabular}{l l l}
\hline
\textbf{Element} & \textbf{Dish Diameter} & \textbf{Gain}\\
\hline
Deep Space Antenna & 35 m & $>$107 dB \citep{dsa_performance} \\
HGA X-band & 3 m & 48 dB \\
MGA S-band & 0.5 m & 20 dB \\
2x UHF on orbiter & - & $>$2 dB \\
\hline
\end{tabular}
\end{table}

\begin{figure}[ht]
\begin{center}
\begin{tabular}{c}
\def\svgwidth{\columnwidth}
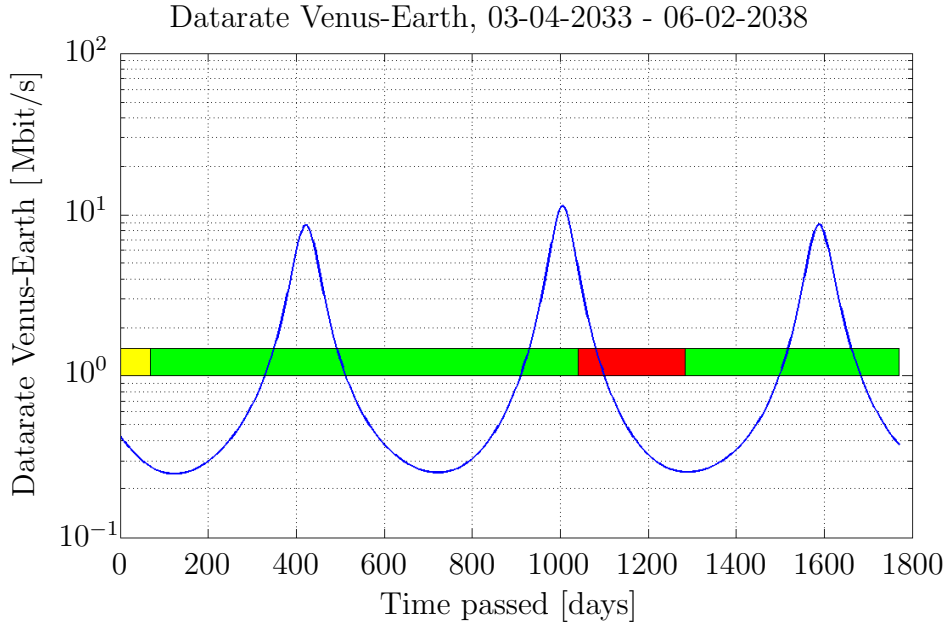
\end{tabular}
\end{center}
\caption 
{ \label{fig:datarate_with_color}
Semi-log plot of data rate varying over mission time. Yellow - Balloon Phase. Green - Science Phase II. Red - Science Phase III, see section~\ref{timeline}.}
\end{figure}

The telecommunication subsystem, including steering equipment and electronics, have an estimated mass of about 68\,kg. Power requirements for the HGA X-band link are about 100\,W, varying with the distance between Venus and Earth. MGA and UHF require just a fraction of this power, and are not listed separately.\\
The X-band link is the most critical one as the scientific data will be transferred to Earth via the HGA. Limiting parameter is the available data rate, which governs operating time and data storage. Figure \ref{fig:datarate_with_color} shows the variation of available data rate in X-band as a function of time and throughout the different science phases, see section \ref{timeline}.

Further design of the communication architecture of the mission should look into using the future European Data Relay System (EDRS) as an Earth-orbit relay, instead of transmitting directly through Earth's atmosphere. Upcoming optical communication technology might be another way of increasing the data rate significantly.\\
The attitude and orbit control system (AOCS) has a similar layout as for the Venus Express \cite{VenusExpress1636}. It consists of three star trackers, two sun acquisition sensors and 2 inertial measurement units for attitude determination. Spacecraft control is provided by five reaction wheels and twelve 10\,N hypergolic bi-propellant attitude control thrusters. The estimated dry mass of the AOCS is about 60\,kg and has a power consumption of about 160\,W.\\
For insertion into Venus orbit a high-thrust engine is necessary. While the propulsion system does not have to bring the orbiter into a circular orbit - which is done using aerobraking instead - it does have to deliver the 1.1\,km/s required to bring the spacecraft into an elliptical orbit around Venus (Venus capture). See for further details section~\ref{timeline}. Apogee motors commonly used for GEO satellites are a feasible option. For simplicity the engine uses the same hypergolic propellant as the AOCS thrusters. Choosing Monomethyl Hydrazine (MMH) and Mixed Oxides of Nitrogen (MON) result in a total propellant mass for engine and AOCS thrusters of about 2100\,kg.\\
The data management system (DMS) is similar to Venus Express as well \cite{VenusExpress1636}. It consists of an on board data handling (OBDH) computer for processing of science and housekeeping date, a solid state mass memory for data dumping and remote terminal units (RTU) forming the interface between the different instruments and the on board computer. The required storage capacity of the mass memory is driven by the data rate of the SAR and the data link capacity to Earth over time. A storage capacity of 20\,Tb is sufficient for the entire mission. The estimated mass of the DMS is about 20\,kg and requires about 70\,W of power.

\begin{figure}[!ht]
\begin{center}
\begin{tabular}{c}
\def\svgwidth{10cm}
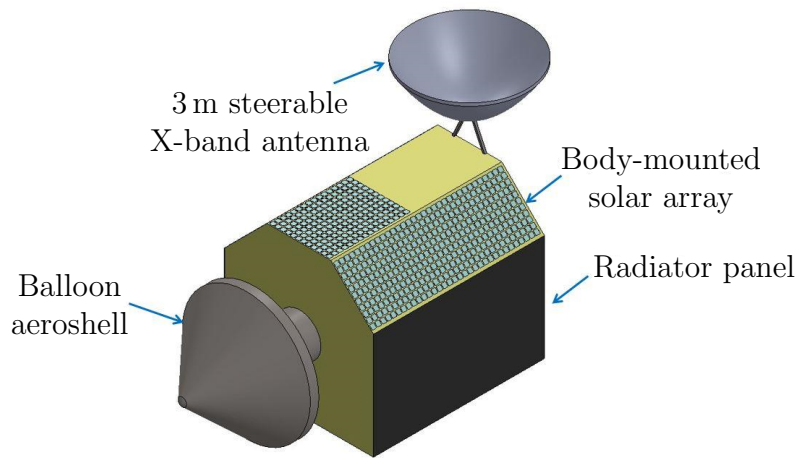
\end{tabular}
\end{center}
\caption 
{ \label{fig:orbiter}
Basic configuration of the orbiter.}
\end{figure} 

The structural architecture of the spacecraft, see figure~\ref{fig:orbiter}, is influenced by multiple factors. The proximity of Venus to the sun ensures that spacecraft orbiting Venus should have no issues generating sufficient power for instrumentation using arrays of solar cells. Having baselined the Ariane 5 as the launch vehicle of choice, it was determined that the payload fairing would have adequate volume to contain a spacecraft of sufficient surface area to accommodate the radiators and allow the use of body mounted solar arrays, not unlike that of CryoSat-2. As was mentioned before, a steerable X-band antenna will be used for the data transfer to Earth. Whilst the addition of such a mechanism is not without the risk associated with moving components, this has been traded against the need to preserve propellant and minimise perturbations incurred by frequent whole spacecraft attitude changes. Furthermore, the use of body mounted solar arrays was determined to be critical in order to avoid perturbing torques on the spacecraft caused by residual atmospheric drag, which would otherwise interfere with the operation of the spacecraft-mounted gradiometer instrumentation. The reduction in the overall drag coefficient of the spacecraft by minimising the frontal area would result in a favourable reduction in propellant required for attitude control, which, in an optimal case, could allow sufficient propellant for a primary mission extension.

\subsection{Balloon}
\label{balloon}
\begin{figure}[!ht]
\begin{center}
\begin{tabular}{c}
\def\svgwidth{10cm}
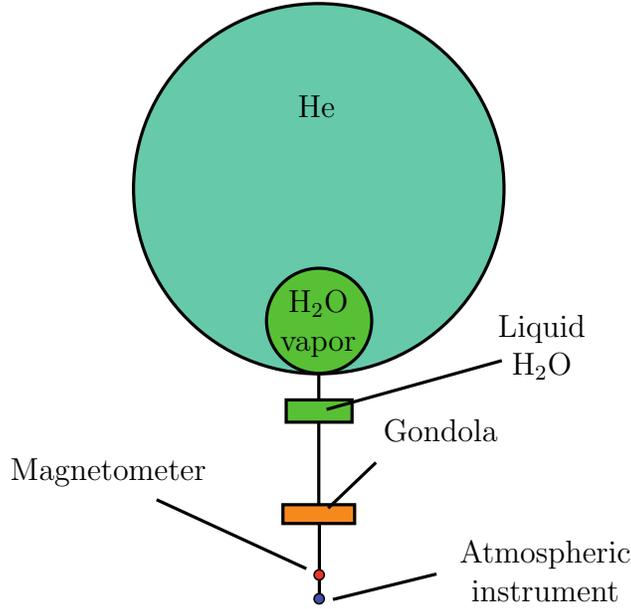
\end{tabular}
\end{center}
\caption 
{ \label{fig:balloon}
Schematic of the dual-phase balloon.}
\end{figure}

A schematic overview of the balloon is given in figure~\ref{fig:balloon}. As shown in this figure the balloon consists of two envelopes, with the smaller one placed in the larger one. The outer shell, which has a diameter of 4.2\,m, contains helium and the inner one, which has a diameter of 1.05\,m, contains water vapour. The inner envelope is connected to a liquid water reservoir. Controlled heat transfer between the liquid water reservoir and Venus' atmosphere results in either condensation of the water vapour in the inner shell or evaporation of water in the reservoir. By increasing or decreasing the amount of water vapour in the inner envelope the buoyancy of the balloon can be changed. With the current design, and based on previous studies by DiCicco et al., the oscillation cycle is estimated to take six hours \cite{DiCicco1995}. This type of balloon is called a phase change balloon with water as the working fluid. However, also ammonia or a combination of the two could be used \cite{Jones1995}.\\
The gondola contains most of the payload and subsystems for command and data handling, communication with the orbiter, thermal control and power provision. A deployable scissors boom is attached to the gondola to which the magnetometers and sounding module are attached. In this way the requirement for both instruments to be as far away as possible from other instruments and structures is met, see section~\ref{payload}.\\
The material of the external envelope must withstand the harsh Venus environment. A possible candidate material is a composite consisting of Teflon-Mylar-Vectran layers. The Teflon outer layer has an excellent resistance to sulphuric acid and the overall envelope structure can withstand the temperature and wind atmospheric conditions \cite{website:VP} \cite{VenusFlagMission}. A full scale test over a period of two weeks has shown that the helium leakage is negligible \cite{Baines2007}.\\
The balloon will be operated at an altitude between 40 and 60\,km. The temperature at this altitude varies between -30$^{\circ}$C to 130$^{\circ}$C. The estimated solar flux is estimated to be about 600 W/m$^2$. The payload requires a temperature between -20$^{\circ}$C up to 50$^{\circ}$C, while the batteries require an operating temperature between 0$^{\circ}$C to 20$^{\circ}$C. To ensure that the temperature stays within the limits a passive thermal control system consisting of MLI and surface coatings and finishes is used. The estimated mass of the thermal control system is about 2\,kg.\\
Power for the payload and other subsystems is provided by primary batteries only. A total of four Lithium-Monofluoride batteries with an energy density of around 300\,Wh/kg \cite{James2003} is used once the balloon has been fully deployed. A Lithium-Sulphur Dioxide battery with an energy density 300\,Wh/kg \cite{James2003} is selected to provide power during the descend phase in the atmosphere, providing power between ejection from the orbiter until the balloon is fully deployed. The mass of the batteries is estimated to be just over 9\,kg, including an uncertainty margin of 5\%.\\
The balloon will be communicating with the orbiter using the UHF-band as was proved practical through the atmosphere on Venus by the Venera landers \cite{Keldych1977}. The communication will utilize two 0.12\,m UHF antennas on the balloon. The whole system has an estimated mass of about 5\,kg and consumes and estimated 10\,W of power.\\
The entry probe consists of a front shell and a back shell, a parachute system to decelerate the probe and deploy the balloon, an adapter to the spacecraft and the balloon itself, formed by the gondola, the aeroshells and the water and helium tanks. The latter will be released once the aeroshell is inflated. Both the front and the back shell are covered by a Carbon Phenolic ablative material layer. This TPS (themal protection system) technology can withstand heat fluxes up to 300\,MW/m2, which is higher than what is expected during entry \cite{Phipps2005}. Additionally, the probe structure and TPS materials are designed to withstand the high deceleration loads ($\sim$60\,g during entry with the configuration described above).\\
The design of the entry probe follows the same philosophy as of several feasibility studies based on the Pioneer Venus heritage: a rigid aeroshell with a medium ballistic coefficient \cite{VenusFlagMission,VenusEP,VenusClimate}. The 45$^{\circ}$ sphere-cone, inspired by the Pioneer Venus entry probes, results in a reduced heat and pressure load on the probe \cite{Dutta2012}.

\subsection{Total mass and power consumption}
\label{massPower}
The total mass of the entry probe is estimated to be about 206\,kg and requires about 103\,W of power. A detailed mass and power breakdown of the different subsystems is provided in table~\ref{tab:balloon_mbudget} and table~\ref{tab:balloon_pbudget}. Indicated in the same table is the contribution of each subsystem to the total mass or power budget. The last column of each table shows the assumed margin, i.e. the quoted number in the second column includes a margin of which the magnitude is quoted in the last column.\\
\begin{table}[!ht]
\setlength\extrarowheight{2pt}
\centering
\caption{Mass budget of the balloon.}
\label{tab:balloon_mbudget}
\begin{tabular}{l c c c}
\hline
\textbf{Subsystem} & \textbf{Mass [kg]} & \textbf{\% of dry mass} & \textbf{Margin [\%]}\\
\hline
Payload & 15.7 & 7.6 & varying \\
Communications & 5.0 & 2.4 & 5 \\
C\&DH/OBDH & 2.1 & 1.0 & 5 \\
Thermal & 1.7 & 0.8 & 10 \\
Power & 8.8 & 4.3 & 5 \\
Structure \& mechanisms & 87.6 & 42.4 & 10 \\
Entry probe & 85.6 & 41.5 & 20 \\\hline
\textbf{total} & \textbf{206.4} & - & - \\
\hline
\end{tabular}
\end{table}
\begin{table}[!ht]
\setlength\extrarowheight{2pt}
\centering
\caption{Power budget of the balloon.}
\label{tab:balloon_pbudget}
\begin{tabular}{l c c c}
\hline
\textbf{Subsystem} & \textbf{Power [W]} & \textbf{\% of total} & \textbf{Margin [\%]}\\
\hline
Payload & 68.7 & 67.0 & varying \\
Communications & 10.5 & 10.2 & 5 \\
C\&DH/OBDH & 5.3 & 5.1 & 5 \\
Thermal & 0.0 & - & - \\
Power & 7.1 & 6.9 & 5 \\
Structure \& mechanisms & 1.1 & 1.1 & 10 \\
Entry probe & 10.0 & 9.7 & 20 \\\hline
\textbf{total} & \textbf{102.6} & - & - \\
\hline
\end{tabular}
\end{table}
The mass of the payload is based the mass of existing instruments. Depending on the required amount of adaptation of existing versions a margin was put on the mass. For instance, the NMS and TLS are well understood and relatively easy to adjust for the proposed mission, so a margin of only 5\% is added to the mass. On the other hand the nephelometer has a lower readiness. For this reason the margin on the mass is set at 30\%. A similar approach was used for estimating the power consumption.\\
The total payload mass, including margins, was used to estimate the mass of the rest of the balloon and gondola. For this purpose the mass of gondola and balloon of the Venus Climate Mission \cite{VCM2010} was scaled according the payload mass. Any uncertainties in the mass estimation is covered by a margin as indicated in the last column of table~\ref{tab:balloon_mbudget}. A similar approach was followed for estimation of the power requirements.\\
The heaviest parts of the entry probe are the front and back shell, including parachute system and S/C adapter, ($\sim$88\,kg) as well as the gondola structure and boom ($\sim$86\,kg). Together, they form more than 84\% of the total entry probe mass. Most of the power, about 2/3, is consumed by the payload.\\
The Hesperos orbiter has an estimated dry mass of about 1490\,kg and requires about 1.5\,kW of power. The mass and power budget of each subsystem is provided in table~\ref{tab:orbiter_mbudget} and table~\ref{tab:orbiter_pbudget}. For each subsystem the mass and power consumption was estimated based on existing hardware sized to the needs of the proposed mission in a similar fashion as for the balloon. Note that the mass and power budget of the AOCS includes the mass and required power of the attitude control thrusters. The tanks in which the propellants are stored are part of the propulsion subsystem.\\
About 50\% of the mass comes from the propulsion system. The payload and power subsystem consume most of the power; each about 35\% of the total.\\

\begin{table}[!ht]
\setlength\extrarowheight{2pt}
\centering
\caption{Mass budget of the orbiter.}
\label{tab:orbiter_mbudget}
\begin{tabular}{l c c c}
\hline
\textbf{Subsystem} & \textbf{Mass [kg]} & \textbf{\% of dry mass} & \textbf{Margin [\%]}\\
\hline
Payload & 271.2 & 18.2 & varying \\
Propulsion & 732.0 & 49.1 & 5 \\
AOCS & 59.6 & 4.0 & 5 \\
Communications & 68.0 & 4.6 & 5 \\
C\&DH/OBDH & 53.8 & 3.6 & 5 \\
Thermal & 5.3 & 0.4 & 5 \\
Power & 50.4 & 3.4 & 5 \\
Structure \& mechanisms & 250.1 & 16.8 & 10 \\\hline
\textbf{total} & \textbf{1490.4} & - & - \\
\hline
\end{tabular}
\end{table}

\begin{table}[!ht]
\setlength\extrarowheight{2pt}
\centering
\caption{Power budget of the orbiter.}
\label{tab:orbiter_pbudget}
\begin{tabular}{l c c c}
\hline
\textbf{Subsystem} & \textbf{Power [W]} & \textbf{\% of total} & \textbf{Margin [\%]}\\
\hline
Payload & 551.4 & 36.4 & varying \\
Propulsion & 50.4 & 3.3 & 5 \\
AOCS & 161.8 & 10.7 & 5 \\
Communications & 106.1 & 7.0 & 5 \\
C\&DH/OBDH & 110.3 & 7.3 & 5 \\
Thermal & 0.0 & - & - \\
Power & 531.0 & 35.0 & 5 \\
Structure \& mechanisms & 4.6 & 0.3 & 10 \\\hline
\textbf{total} & \textbf{1515.5} & - & - \\
\hline
\end{tabular}
\end{table}

\section{Mission Phases and Timeline}
\label{timeline}
Given the dry mass of the orbiter and entry probe of about 1700\,kg in total while keeping in mind the propellant mass needed, an Ariane 5 rocket is required for launch. A suitable launch window will open in December 2032. After launch the satellite is sent on an interplanetary Hohmann transfer trajectory to Venus. The transfer will take about 117 days. A burn with an apogee kick motor will bring the orbiter with entry probe in an elliptic orbit around Venus with an inclination of 85$^{\circ}$ and a periapsis and apoapsis of 250\,km and 19500\,km, respectively. In this orbit, some operations will take place before the final circular orbit is achieved, this is called the Balloon Phase. The balloon will be deployed in this phase in order to reduce the mass before the transference to a circular orbit and hence, reduce the propellant mass requirements. Once this phase is completed, an aerobraking manoeuvre will take place to circularise the orbit. This manoeuvre consists on taking advantage of the drag from the planet's atmosphere at the perigee in order to reduce the apogee altitude with very low fuel requirements until the final desired altitude is reached. When this altitude is achieved, another $\Delta$V is given in order to enter and maintain the final circular orbit and once in this orbit, the Orbiter Phase will begin then.

\subsection{Balloon phase}
\label{balloonPhase}
The balloon is deployed as soon as the spacecraft is in its initial elliptical orbit and before the aerobraking manoeuvre. The entry probe will be released between the apoapsis and the periapsis from a highly elliptical orbit to reduce the entry velocity to about 8.6\,km/s. Given the kinetic energy, a steep entry flight path angles (EFPAs) of $\sim$40$^{\circ}$ can be used, while maintaining the g-load within acceptable limits.\\
The correct attitude of the probe is crucial for the TPS to work as designed. To ensure static stability, the probe will be spin stabilised \cite{Lorenz2006}. The spin rate and the required velocity will be delivered by the spacecraft and thus reducing the complexity and weight of the entry probe \cite{Phipps2005,VenusEP}. After release of the entry probe a small course correction of the orbiter is required to return it to its initial orbit.\\
Once the probe reaches the outer edge of the atmosphere, assumed at an altitude of 200\,km, the induced drag causes rapid deceleration. At an altitude of 70\,km the front and back shell are separated by a pyrotechnic mechanism. The front shell will under influence of gravity accelerate towards the surface. The back shell acts as a drogue chute and pulls the main parachute for further  deceleration of the probe. The parachute itself pulls the balloon, after which inflation of both balloon envelopes start. Once the balloon envelopes are fully inflated, both the parachute and the main helium tank are released. At this point the balloon is at an altitude of about 60\,km and checkout and commissioning of all systems will start.\\
As was explained in section~\ref{balloon} the balloon will be oscillating between 40 and 60\,km above Venus' surface in a period of about six hours. Given the energy available from the batteries and the power consumption of the payload and subsystems the balloon phase will last for about 25 days. Due to prevailing wind directions on Venus, the balloon will circumnavigate the planet a couple of times before the batteries run out. A secondary effect is the drift towards one of the poles during this time. As such, not only a vertical profile of Venus' cloud-level atmosphere will be obtained but also as a function of latitude.\\
The mass spectrometer measurements are considered as most important to understand the general properties of Venus' atmosphere, thus they are conducted at the beginning of the balloon phase. As they are very power intensive only three measurements at highest balloon position and three measurements at lowest balloon position are foreseen. The other balloon instruments need far less power and are alternately operated in a continuous loop interrupted only for the spectrometer measurements. During one loop the magnetometer and sounding module collect data for 20\% of the time each and the nephelometer for 60\% of the time.\\
During the balloon release and science phase the orbiter is in a highly elliptical orbit and acting as relay station for the science data returned by the balloon. At the same time the balloon will be followed by the IR and UV spectrometers on board the orbiter. This is designated as phase I of the orbiter.\\
The end of the balloon phase is scheduled for May 2033. Shortly before the energy supply fails the gondola will be cut off. It will be in free fall for approximately 2 minutes before it hits the ground. During this time the magnetometer is switched on. As the measurement cycle of this instrument is very short it is possible to obtain data during this short period.

\subsection{Orbiter phase}
\label{orbSciPhase}
After finishing the balloon phase the orbiter will be put in a circular orbit around Venus, which will be achieved by aerobraking in the upper atmosphere of Venus as mentioned before. This technique has been successfully demonstrated by Venus Express \citep{VALSERRA2011}. The final orbit will have an altitude of 250\,km with an inclination of 85$^{\circ}$. The orbital parameters have been chosen such that a repeat track is achieved every two Venus days. After an initial checkout and commissioning phase the main science phase starts in June 2033 with a nominal duration of seven Venus days or about five Earth years. The nominal mission ends in February 2038. Although 5 years is a long period for the primary science phase of a planetary mission, experience with other missions shows that such long durations are not uncommon. For instance, the primary science phase of ESA’s Venus Express mission was planned for 2 years, but the mission was extended 5 times, lasting over 9 years \citep{Svedhem2007,Svedhem2009}. Also this long duration would allow to tune the observations according to the requirements and constraints that would arise once the missions is operative. For instance, the extension of the Venus Express missions had the main objective of adding new fine-tuned observations from lessons learned over the ongoing mission that can only arise once the mission is operative \citep{Hoofs2009}. Depending on the resources left the Hesperos mission can also be extended.\\
The main science phase of the orbiter consists of two phases. Phase II is dedicated to surface mapping and uses the SAR as well as the IR and UV spectrometers/cameras. Phase III is dedicated to gravity gradient mapping. During this phase only the cold atom gradiometer will be operated.\\
The science mission of the orbiter starts with Phase II which lasts for four Venus days. To prevent the SAR from overheating it will be operated for half an orbit and then switched off to cool down. To ensure effective cooling, the SAR is operated on the day side and switched off on the night side of Venus. While cooling down, IR spectrometers/cameras are switched on. Note that the IR camera can only be used on the night side and as such is a good combination for the SAR.\\
During the first two Venus days of phase II the whole surface is mapped with the SAR. This first sweep of the surface is used as reference. A second sweep, lasting two Venus days again, is required to establish a topographical map. The end of Phase II is on the night side of Venus with spectrometers/cameras switched on.\\
Then Phase III starts which is fully dedicated to gravity gradient measurements with the cold atom gradiometer. This instrument requires a disturbance free environment, i.e. any noise in the form of sound, electric current or temperature above the background temperature should be avoided. For this reason, Phase II should end at the night side of Venus so the SAR has time to cool down sufficiently before the gradiometer is switched on. Phase III ends after one Venus day.\\
After Phase III another radar mapping of the Venusian surface takes place, but this time for two Venus days only. This partial repetition of Phase II is to detect possible changes in the topography. Here the first full surface scan during the first two Venus days of Phase II is used as reference again. The obtained topographical map can then be compared with the topographical map established earlier and will show changes in topography over the course of one full Venus day.\\
At several moments during the mission thruster firings are required for orbit insertion and orbit maintenance. The required $\Delta$V for each type of manoeuvre is shown in table~\ref{deltaV}. The numbers include a margin used for calculating the required propellant mass.

\begin{table}[!ht]
\setlength\extrarowheight{2pt}
\centering
\caption{$\Delta$V budget}
\label{deltaV}
\begin{tabular}{lc}
\hline
\textbf{Manoeuvre} & \textbf{$\Delta$V (km/s)} \\ \hline
Hohmann Transfer & 2.84 \\
Elliptical Orbit & 1.47 \\
Aerobreaking     & 0.001 \\
Circular Orbit   & 0.001 \\
Maintenance      & 0.11 \\ \hline
\textbf{Total}   & \textbf{4.83} \\ \hline
\end{tabular}
\end{table}

Given the dry mass of the satellite, as mentioned in the previous paragraph, the propellant mass can be calculated. A margin of 20\% is used for dry mass. It is further assumed that the specific impulse, I$_{sp}$, of the apogee kick motor is 321\,s and of the attitude control thrusters 291\,s. Combined with the $\Delta$V requirements as shown in table \ref{deltaV} this results in a propellant mass of just over 2100\,kg. The total spacecraft mass is thus almost 4100 kg.

\begin{table}[!ht]
\setlength\extrarowheight{2pt}
\centering
\caption{Overall mass budget}
\label{totalmbudget}
\begin{tabular}{l c}
\hline
 & \textbf{Mass [kg]} \\ \hline
Balloon total & 206.4 \\
Orbiter total & 1490.4 \\
S/C dry mass + 20\% margin &  1978.8 \\
Propellant mass &  2101.6 \\
Launch mass &  4080.4 \\\hline
\end{tabular}
\end{table}

\section{Conclusions}
\label{conclusions}
A comprehensive understanding of Venus, its evolution and why it is so different from Earth, despite a couple of striking resemblances, is still lacking. Clarifying some of the open questions not only helps understanding how Venus evolved to the planet as we know it, but will also help in constraining planetary models in general and help understand and interpret the variety of exoplanets discovered in the past decade.\\
In order to obtain this understanding investigating the inner structure and the atmosphere as well as the interaction with each other need to be studied. To achieve this, monitoring of the planet from orbit as well as in-situ measurements are required. The required technology for such mission is for most part present already. However, a couple of instruments and subsystems require further development before it can be employed as envisaged for the mission. These includes the balloon and gondola as well as the cold atom gravity gradiometer.

\section*{Acknowledgement}
\label{acknowledge}
The authors would like to express their gratitude to FFG and ESA for organising the Alpbach Summer School as well as the post-Alpbach week. We would also like to thank our supervisors G\"{u}nter Kargl and Olivier Baur from the Space Research Institute of the the Austrian Academy of Sciences, Richard Ghail from Imperial College London and Manuela Unterberger from the Technical University of Graz for their valuable advice and enthusiasm. The authors are also greatful for the many useful comments from Colin Wilson during the review process. Author A.\,\L{}osiak would further like to acknowledge the NCN grant, which made the participation in this study possible (2013/08/S/ST10/00586).

\bibliographystyle{model1-num-names}
\bibliography{refs_submission.bib}

\end{document}